\documentclass[prd,aps,footinbib,twocolumn,notitlepage,longbibliography,unsortedaddress,superscriptaddress]{revtex4-1}
\usepackage{bm}
\usepackage{amsmath}
\usepackage{amssymb}
\usepackage{graphicx}
\usepackage{slashed}
\usepackage{accents}
\usepackage{mathrsfs}
\usepackage{braket}
\usepackage{units}

\usepackage{upgreek}
\usepackage{booktabs}
\usepackage{multirow}
\usepackage{minibox}

\DeclareSymbolFont{symbolsC}{U}{txsyc}{m}{n}
\SetSymbolFont{symbolsC}{bold}{U}{txsyc}{bx}{n}
\DeclareFontSubstitution{U}{txsyc}{m}{n}

\AtBeginDocument{
\heavyrulewidth=.08em
\lightrulewidth=.05em
\cmidrulewidth=.03em
\belowrulesep=.65ex
\belowbottomsep=0pt
\aboverulesep=.4ex
\abovetopsep=0pt
\cmidrulesep=\doublerulesep
\cmidrulekern=.5em
\defaultaddspace=.5em
\tabcolsep=4pt
}

\usepackage{xcolor}
\newcommand{\hl}[1]{{\color{black}#1}}

\usepackage[breaklinks=true,colorlinks=true,linkcolor=blue, citecolor=blue, urlcolor=blue]{hyperref}
\usepackage{breakurl}
\usepackage{minibox}
\DeclareSymbolFont{eulerscript}{U}{eur}{m}{n}
\DeclareSymbolFontAlphabet{\matheuler}{eulerscript}
\DeclareMathAlphabet{\boldgreek}{OML}{zplm}{b}{it}
\newcommand{\nfrac}[2]{{#1}/{#2}}
\newcommand{\eps}{\epsilon}
\newcommand{\mc}[1]{\mathcal{#1}}

\newcommand{\sdist}{\kern 0.20em}
\renewcommand{\eqref}[1]{Eq.\sdist(\ref{#1})}
\newcommand{\figref}[1]{Fig.\sdist\ref{#1}}
\newcommand{\tabref}[1]{Tab.\sdist\ref{#1}}

\begin{document}

\title{On the Prospect of Studying Nonperturbative QED with Beam-Beam Collisions}

\author{V.~\surname{Yakimenko}}
\email{yakimenk@slac.stanford.edu}
\affiliation{SLAC National Accelerator Laboratory, Menlo Park, CA USA}
\author{S.~\surname{Meuren}}
\affiliation{Department of Astrophysical Sciences, Princeton University, Princeton, NJ USA}
\author{F.~Del Gaudio}
\affiliation{GoLP/Instituto de Plasmas e Fus\~ao Nuclear, Instituto Superior T\'ecnico, Universidade de Lisboa, Lisboa, Portugal}
\author{C.~\surname{Baumann}}
\affiliation{Heinrich-Heine-Universit\"at, D\"usseldorf, Germany}
\author{A.~\surname{Fedotov}}
\affiliation{National Research Nuclear University MEPhI, Moscow, Russia}
\author{F.~\surname{Fiuza}}
\affiliation{SLAC National Accelerator Laboratory, Menlo Park, CA USA}
\author{T.~\surname{Grismayer}}
\affiliation{GoLP/Instituto de Plasmas e Fus\~ao Nuclear, Instituto Superior T\'ecnico, Universidade de Lisboa, Lisboa, Portugal}
\author{M.\,J.~\surname{Hogan}}
\affiliation{SLAC National Accelerator Laboratory, Menlo Park, CA USA}
\author{A.~\surname{Pukhov}}
\affiliation{Heinrich-Heine-Universit\"at, D\"usseldorf, Germany}
\author{L.\,O.~\surname{Silva}}
\affiliation{GoLP/Instituto de Plasmas e Fus\~ao Nuclear, Instituto Superior T\'ecnico, Universidade de Lisboa, Lisboa, Portugal}
\author{G.~\surname{White}}
\affiliation{SLAC National Accelerator Laboratory, Menlo Park, CA USA}

\begin{abstract}
% possibility -> experimental feasibility
% I moved "with a 100 GeV-class particle collider", as the statement "By using tightly compressed and focused electron beams, beamstrahlung radiation losses can be mitigated" is not only true for 100 GeV class colliders. I think.
%
% Do we need the "for the first time" here?
%
 \hl{We demonstrate the experimental feasibility of probing the fully nonperturbative regime of quantum electrodynamics with a 100\,GeV-class particle collider}. By using tightly compressed and focused electron beams, beamstrahlung radiation losses can be mitigated, allowing the particles to experience extreme electromagnetic fields. Three-dimensional particle-in-cell simulations confirm the viability of this approach. The experimental forefront envisaged has the potential to establish a novel research field and to stimulate the development of a new theoretical methodology for this yet unexplored regime of strong-field quantum electrodynamics. 
\end{abstract}

\date{\today}

\maketitle

The interaction of light and matter is governed by quantum electrodynamics (QED), which
is the most successfully tested theory in physics. According to the present understanding of QED, the properties of matter change qualitatively in the presence of strong electromagnetic fields. The importance of strong-field quantum effects is determined by the Lorentz invariant parameter $\chi = E^*/E_{\mathrm{cr}}$ \cite{landau_quantum_1981,di_piazza_extremely_2012} (also called beamstrahlung parameter in the context of particle colliders), which compares the electromagnetic field in the electron/positron rest frame $E^*$  with the QED critical field %
$E_{\mathrm{cr}} =\nfrac{m^2 c^3}{(e \hbar)} \approx\unitfrac[1.32 \times 10^{18}]{V}{m}$. Here, $m$ and $e$ are the electron/positron mass and charge, $c$ is speed of light, and $\hbar$ is reduced Planck constant, respectively.
%
%$E_{\mathrm{cr}} =\frac{m^2 c^3}{c \hbar} \approx \unitfrac[1.32 \times 10^{18}]{V}{m}$. Here $m$ and $e$ are mass and charge of electron/positron, $c$ is speed of light, and $\hbar$ is reduced Planck constant. 
%
Whereas classical electrodynamics is valid if $\chi\ll 1$, quantum effects like the recoil of emitted photons (quantum radiation reaction) and the creation of matter from pure light become important in the regime $\chi \gtrsim 1$. Eventually, the interaction between light and matter becomes fully nonperturbative if $\chi \gg 1$.

{The behavior of matter near QED critical field strengths (i.e., the regime $\chi \sim 1$) is important} in astrophysics (e.g., gamma-ray bursts, pulsar magnetosphere, supernova explosions) \cite{harding_physics_1991,uzdensky_plasma_2014,cerutti_electrodynamics_2017}, \hl{at the interaction point of future linear particle colliders \cite{yokoya_beam-beam_1992,chen_coherent_1989,baier_quantum_1989,bell_quantum_1988,chen_introduction_1988,noble_beamstrahlung_1987,blankenbecler_quantum_1987,jacob_quantum_1987}}, and in upcoming high energy density physics experiments, where laser-plasma interactions will probe quantum effects \cite{eliwhitebook}. Experimental investigations of strong-field QED  have just approached $\chi\lesssim 1$, e.g., by combining highly energetic particles with intense optical laser fields. This experimental scheme, first realized in the SLAC E-144 experiment \cite{bula_observation_1996,burke_positron_1997}, has been recently revisited \cite{cole_experimental_2018,poder_evidence_2017}. Notable alternatives are x-ray free electron lasers \cite{cartlidge_light_2018}, highly charged ions \cite{ullmann_high_2017}, heavy-ion collisions \cite{rafelski_probing_2017}, and strong crystalline fields \cite{wistisen_experimental_2018}. {The success of QED in the regime $\chi \lesssim 1$ is based on the smallness of the fine-structure constant $\alpha \approx 1/137$, which facilitates perturbative calculations.} 

{Inside an extremely strong electromagnetic background field, however, the situation changes profoundly. According to the Ritus-Narozhny conjecture the actual expansion parameter of QED in the strong-field sector $\chi \gg 1$ is $\alpha\chi^{2/3}$. \cite{ritus_radiative_1972,narozhny_expansion_1980,fedotov_conjecture_2017}. Correspondingly, QED becomes a strongly coupled theory if $\alpha\chi^{2/3} \gtrsim 1$ and the so-called dressed loop expansion breaks down. This implies that the emission of a virtual photon by an electron/positron or the temporarily conversion of a photon into a virtual electron-positron pair is no longer an unlikely event. {Therefore, the existing theoretical framework is not suitable for this regime.}

{The fully nonperturbative sector ($\alpha\chi^{2/3} \gtrsim 1$) is currently seen as beyond experimental reach.} The fundamental challenge in probing such extreme fields is the fast radiative energy loss by electrons/positrons. Its mitigation requires the switching time of the background field to be smaller than the electron/positron radiative life time { $\tau_l \sim \gamma \tau_c/(\alpha\chi^{2/3})$ ($\tau_c = \lambdabar_c/c \approx \unit[1.3 \times 10^{-21}]{s}$; $\lambdabar_c =\hbar / (m c) \approx \unit[3.9 \times 10^{-13}]{m}$ and $\gamma$ denotes the Lorentz gamma factor) \cite{meuren_quantum_2011}. \hl{As the spatial extend of an optical laser pulse must be at least of the order of the laser wavelength $\lambda_l \sim \upmu m$, we need a multiple TeV electron/positron beam ($\gamma \sim 10^7$) to ensure $\lambda_l  \lesssim \gamma \: \lambdabar_c$. Therefore, reaching the regime $\alpha\chi^{2/3} \gtrsim 1$ with electron-laser collisions is not viable at the 100 GeV scale.}

\begin{figure*}
	\centering
	\includegraphics[width=0.8\textwidth]{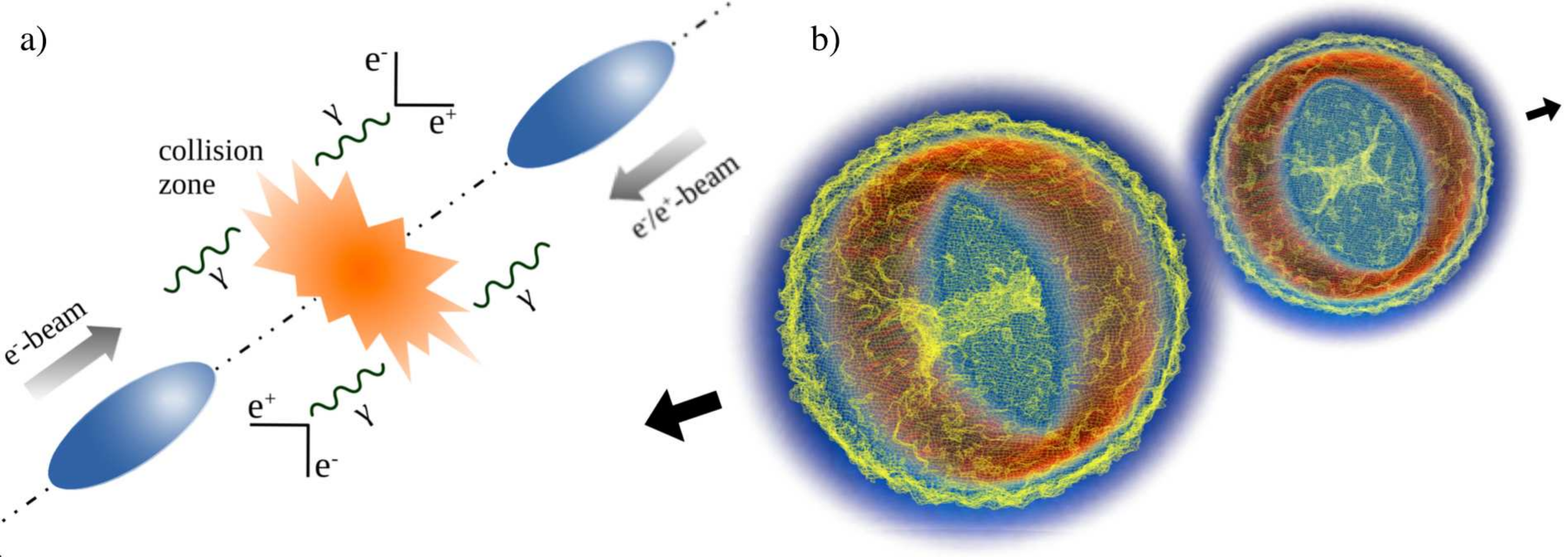} 
	\caption{\label{fig:collider} a) Illustration of a beam-beam collider for probing the fully nonperturbative QED regime. b) 3D OSIRIS-QED simulation of the collision of two spherical $\unit[10]{nm}$ electron beams with $\unit[125]{GeV}$ energy (blue). The fully nonperturbative QED regime $\alpha\chi^{2/3} \geq 1$ is experienced by 38\% of the colliding particles (red). The interaction produces two dense gamma-ray beams with $0.2$ photons with $E_\gamma \geq 2 mc^2$ per primary electron (yellow).}
\end{figure*}

In this Letter, we show that using tightly compressed and focused beams it is possible to probe for the first time the fully nonperturbative QED regime with a 100 GeV-class particle collider (\figref{fig:collider}). We argue that these beams could be produced with accessible technology. Full 3D particle-in-cell (PIC) simulations confirm the possibility of limiting beam energy losses to $\lesssim 5\%$, \hl{implying that the majority of particles reach the strong field region}. 

 \hl{To estimate the importance of nonperturbative effects, we take phenomenologically into account that quantum fluctuations dynamically increase the effective electron/positron mass and thus the effective QED critical field. As a result, one expects that radiation and pair production are attenuated with respect to the perturbative predictions. Our simulations show that corrections on the order of $20-30\%$ are to be expected (see below). Correspondingly, nonperturbative effects should be observable with a 100\,GeV-class particle collider.}

The breakdown of perturbation theory in the regime $\alpha\chi^{2/3} \gtrsim 1$ has an intuitive explanation. In vacuum, the characteristic scales of QED are determined by the electron/positron mass $m$. In the presence of a background field, however, the fundamental properties of electrons, positrons, and photons are modified by quantum fluctuations (\figref{fig:radcorrections}). Figuratively speaking, the quantum vacuum is not empty but filled with virtual electron-positron pairs. A strong electromagnetic field polarizes/ionizes the vacuum, which therefore behaves like an electron-positron pair plasma. As a result, the ``plasma frequency of the vacuum'' changes the photon dispersion relation, implying that a photon acquires an effective mass $m_\gamma(\chi)$, see Supplemental Material. The appearance of a photon mass induces qualitatively new phenomena like vacuum birefringence and dichroism \cite{bragin_high-energy_2017,nakamiya_probing_2017,king_vacuum_birefringence_2016,ilderton_prospects_2016}. Perturbation theory is expected to break down in the regime $m_\gamma(\chi) \gtrsim m$, where modifications due to quantum fluctuations become of the same order as the leading-order tree-level result (\figref{fig:radcorrections}). 

\hl{In order to provide an intuitive understanding for the scaling of $m_\gamma(\chi)$}, a photon with energy $\hbar\omega_\gamma \gg mc^2$ is considered, which propagates through a perpendicular electric field with magnitude $E$ in the laboratory frame. The $\chi$ associated with this photon is $\chi \sim \gamma E/E_{\mathrm{cr}}$, where $\gamma = \hbar\omega_\gamma/(mc^2)$ can be interpreted as a generalized Lorentz gamma factor. As the polarization of the quantum vacuum requires at least two interactions (\figref{fig:radcorrections}), it is expected that $m_\gamma^2(\chi) \sim \alpha M^2$ (the plasma frequency of a medium exhibits the same scaling in $\alpha$). Here, $M \sim eE\Delta t/c$ denotes the characteristic mass scale induced by the background field and $\Delta t$ represents the characteristic lifetime of a virtual pair.

\begin{figure}
	\centering
	\includegraphics[width=0.45\textwidth]{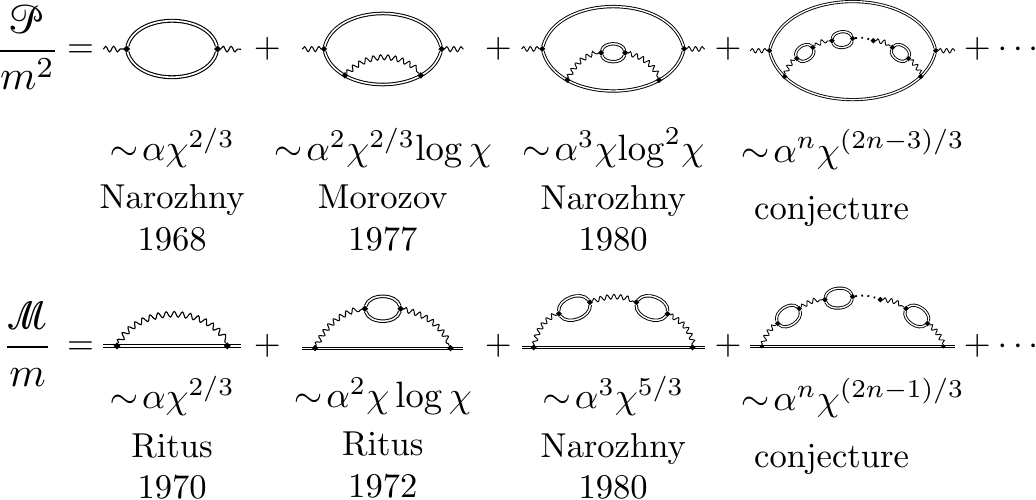} 
	\caption{\label{fig:radcorrections} Dressed loop expansion of the polarization operator $\mc{P}$ (top row) and mass operator $\mc{M}$ (bottom row). Wiggly lines denote photons and double lines dressed electron/positron propagators \cite{di_piazza_extremely_2012}. According to the Ritus-Narozhny conjecture, the diagrams shown represent the dominant contribution at n-loop and $\alpha\chi^{2/3}$ is the true expansion parameter of strong-field QED in the regime $\chi \gg 1$  \cite{ritus_radiative_1972,narozhny_expansion_1980,fedotov_conjecture_2017}.}
\end{figure}

\hl{The scaling of $\Delta t$ is determined by the Heisenberg uncertainty principle $\Delta t \Delta \eps \sim \hbar$, where $\Delta\eps \!=\! \eps_- \!+\! \eps_+ \!-\! \eps_\gamma$ quantifies energy non-conservation at the pair production vertex. Here, $\eps_- \approx \eps_+ = \sqrt{(pc)^2 + m^2c^4 + (eE \Delta t c)^2} \approx pc + (eE\Delta t c)^2/ (2pc)$ are the electron/positron energies and $\eps_\gamma = p_\gamma c$ is the energy of the gamma photon (electron and positron have the same initial momentum $p = p_\gamma/2$ at threshold). Assuming, $\chi\gg 1$ and thus $eE\Delta t \gg mc$ (momentum acquired by the charges in the background field $E$), we find $\Delta\eps \sim (eE\Delta t c)^2 / (\hbar\omega_\gamma)^2$. }
Notably, the resulting field-induced mass scale $M \sim eE\Delta t/c \sim m\chi^{1/3}$ is independent of $m$ (note that $\chi \sim m^{-3}$). This suggests a new regime of light-matter interaction, where the characteristic scales of the theory are determined by the background field ($M \gg m$). The scaling $m_\gamma^2(\chi) \sim \alpha M^2 \sim \alpha \chi^{2/3} m^2$ in the regime $\chi\gg 1$ implies $m_\gamma \gtrsim m$ if $\alpha\chi^{2/3} \gtrsim 1$ and thus a breakdown of perturbation theory at the conjectured scale \cite{ritus_radiative_1972,narozhny_expansion_1980,fedotov_conjecture_2017}. The same scaling is also found for the electron/positron effective mass by analyzing the mass operator (see Supplemental Material). 

\hl{A similar breakdown of perturbation theory is predicted for supercritical magnetic fields [$B \gg B_{cr} = m^2 c^3/(e \hbar) \approx \unit[4.41 \times 10^9]{T}$]. Whereas the mass correction for electrons in the lowest Landau level scales logarithmically \cite{jancovici_radiative_1969}, photons acquire an effective mass via the polarization operator, which exhibits a power-law scaling \cite{shabad_photon_2004,shabad_photon_1975} (for a discussion of possible astrophysical observables see, e.g., \cite{shabad_-quanta_1982}). For supercritical magnetic fields effective dimensional reduction facilitates nonperturbative calculations \cite{gusynin_dynamical_1999,gusynin_electron_1999,loskutov_behavior_1981}. Note that the case considered here is complementary and qualitatively different, as it corresponds on the contrary to ultrarelativistic electrons/positrons occupying very high Landau levels. As a result, they can emit photons and produce pairs and thus provide two accessible observables which are affected by radiative corrections.}

The key challenge for reaching the fully nonperturbative regime $\alpha\chi^{2/3} \gtrsim 1$ in beam-beam collisions is the mitigation of radiative losses through beamstrahlung: the emission of radiation as the colliding particles are bent in the fields of the opposing bunch. This process is characterized by four beam parameters: the transverse $\sigma_r$ and the longitudinal $\sigma_z$ dimensions of the bunches ($\sigma_r = \sigma_x = \sigma_y$ for radially symmetric beams), the number of particles per bunch $N$ (i.e., the total charge) and the beam Lorentz factor $\gamma$. Lorentz invariance requires that only the ratio $\sigma_z^* = \sigma_z / \gamma$ is relevant, implying three independent degrees of freedom. 

The total radiation probability $W$ (per beam particle) and the
disruption parameter $D$, which characterizes the transverse motion of
the beam particles, scale as
\begin{gather}
W \!\sim\! \alpha \chi^{2/3}_{\mathrm{av}} \, \frac{\sigma_z^*}{\lambdabar_c},
\,\,
D \!\sim\! \frac{N \alpha \lambdabar_c \sigma^*_z}{\sigma_r^2},
\,\,
\chi_{\mathrm{av}} \!\approx\! \frac{5}{12} \frac{N \alpha
\lambdabar^2_c}{\sigma_r\sigma^*_z},
\end{gather}
where $\chi_{\mathrm{av}}$ denotes the average value of the
beamstrahlung parameter $\chi$ (in the accelerator science literature
the symbol $\Upsilon=\chi_{\mathrm{av}}$ is commonly used). The given
estimate for $\chi_{\mathrm{av}}$ holds for a radially symmetric
Gaussian charge density profile \cite{chen_coherent_1989}. In order to
achieve a controlled interaction $D \ll 1$ is desirable, which implies
that the classical trajectories of the colliding particles are only
slightly distorted.

\begin{table}
\caption{\label{tab:parameters} Comparison between the parameters of the Nonperturbative QED (NpQED) collider discussed here and other existing linear accelerator/collider designs. Collision parameters for FACET-II \cite{FACETIITDR} are not applicable, as it has only one beam. Here $\chi_{\mathrm{av}}$ and $\chi_{\mathrm{max}}$ for ILC \cite{ILC} and CLIC \cite{CLIC} are calculated without taking into account the expected change in the beam size during collision, which is characteristic for high disruption parameters.}

\begin{center}
\begin{tabular}{l@{\hskip -5pt}ccccc}
\toprule
\rotatebox{90}{\parbox[c]{1.5cm}{\rotatebox{-90}{Parameter}}} & \rotatebox{90}{\parbox[c]{1.5cm}{[Unit]}} & \rotatebox{90}{\parbox[c]{1.5cm}{\minibox[c]{NpQED\\ Collider}}} & \rotatebox{90}{\parbox[c]{1.5cm}{\minibox[c]{FACET-II}}} & \rotatebox{90}{\parbox[c]{1.5cm}{\minibox[c]{ILC}}} & \rotatebox{90}{\parbox[c]{1.5cm}{\minibox[c]{CLIC}}}\\
\midrule

Beam Energy  & [GeV] 					&	125			&	10		&	250			&	1500		\\
Bunch Charge & [nC] 					&	1.4			&	1.2		&	3.2			&	0.6			\\
Peak Current & [kA] 					&	1700		&	300		&	1.3			&	12.1		\\
Energy Spread (rms) &[\%]				&	0.1			&	0.85	&	0.12		&	0.34		\\
Bunch Length (rms)& [$\unit{\upmu m}$]	&	0.1-0.01		&	0.48	&	300			&	44			\\[3pt]
Bunch Size (rms)& [$\unit{\upmu m}$]	&	\minibox[c]{0.01\\0.01}	&	\minibox[c]{3\\2}		&	\minibox[c]{0.47\\0.006}	&	\minibox[c]{0.045\\0.001}	\\[8pt]
\minibox[l]{Pulse Rate $\times$\\ Bunches/Pulse} & \minibox[c]{[Hz]$\times$ \\ $N_{\mathrm{bunch}}$} & \minibox[c]{100$\times$\\1} & \minibox[c]{30$\times$\\1} & \minibox[c]{5$\times$\\1312} & \minibox[c]{50$\times$\\312} \\[8pt]
\multirow{2}{*}{\minibox[l]{Beamstrahlung\\ Parameter}} & $\chi_{\text{av}}$		& 969	& --		&	0.06	&	5		\\
 & $\chi_{\text{max}}$		& 1721	& --		&	0.15	&	12		\\[3pt]
\multirow{2}{*}{\minibox[l]{Disruption\\ Parameters}} & \multirow{2}{*}{$D_{x,y}$}		& 0.001	& --		&	0.3	   &	0.15	\\
 & 																						& 0.001	& --		&	24.4   &	6.8		\\[3pt]
Peak electric field & [TV/m] & $4500$ & $3.2$ & $0.2$ & $2.7$ \\ 
Beam Power			& [MW] & $10^{-3}$ & $10^{-4}$ & 5 & 14 \\
Luminosity			& [$\unit{cm^{-2}s^{-1}}]$ & $10^{30}$ & -- & $10^{34}$ & $10^{34}$\\
\bottomrule		
\end{tabular}
\end{center}
\end{table}

The requirements given above ($\alpha\chi^{2/3} \gtrsim 1$, $D \lesssim 0.01$, and $W \lesssim 1$) constrain the three beam parameters: $N \gtrsim 1/\alpha^4 \sim 10^9$ (i.e., $\gtrsim \unit[0.1]{nC}$ per bunch), $\sigma_r \sim 10 \sqrt{N\alpha} \lambdabar_c \sim \unit[10]{nm}$, and $\sigma_z^* \lesssim \lambdabar_c$. For a beam energy of $\approx \unit[100]{GeV}$ ($\gamma \approx 2 \times 10^5$) this implies $\sigma_z \lesssim \unit[100]{nm}$. In general, decreasing $\sigma_z$ is beneficial for all three parameters ($\chi$, $D$, $W$), whereas increasing the charge must be accompanied by a transverse compression to keep the disruption parameter small. According to these considerations, the natural set of parameters for a $\sim\unit[100]{GeV}$ nonperturbative QED (NpQED) collider, which is capable of reaching $\alpha\chi^{2/3} \gtrsim 1$ with low disruption, is given in \tabref{tab:parameters}.

The NpQED collider discussed here maximizes the beam fields by employing highly compressed and round bunches. This approach differs significantly from existing linear collider designs like ILC \cite{ILC} or CLIC \cite{CLIC}, which use flat ($\sigma_x/\sigma_y > 10$) beam configurations to avoid strong fields and optimize the luminosity. The idea for this NpQED collider originated from SLAC's FACET-II \cite{FACETIITDR}, designed to generate beams with up to 300 kA peak current ($\sigma_z \sim \unit[0.5]{\upmu m}$) at 10 GeV energy. Merging the high energy, high transverse quality beams of linear collider designs with the high peak compression of FACET-II encapsulates the key design challenges (\tabref{tab:parameters}).

Nonperturbative QED can be probed with either electron-electron or electron-positron collisions. Using only electrons is preferable, as it avoids the challenge of generating positrons with the required longitudinal brightness. Next-generation cryogenic photoinjectors \cite{rosenzweig_next_2016} aim for a factor $>4$ improvement in emittance ($\sim \unit[35]{nm-rad}$ at $\unit[100]{pC}$). This will translate into electron focusing requirements similar to CLIC.  

In order to obtain a compact accelerator design, high gradient technology (e.g., X-band radio frequency or plasma-based acceleration) could be employed, leading to an accelerating section with a footprint comparable with the SLAC linac. The fully nonperturbative QED regime can be studied with a single bunch per pulse at $\sim$ 100 Hz. An ILC-type linac and repetition rate might be required for compression stability and feedback systems. This would result in a luminosity equivalent to ILC at a much lower beam power.

The required bunch compression extends the state-of-the-art FACET-II design by a factor of $5$. The anticipated increase of collective effects during bunch compression can be compensated by using advanced mitigation strategies, e.g., based on Coherent Synchrotron Radiation (CSR) suppression and/or shielding techniques \cite{yakimenko_experimental_2012,jing_compensating_2013}.
 
The final focus system can be based on the CLIC design, as the requirements are similar. However, delivering round beams with the required chromaticity compensation presents a unique challenge, especially when coherent effects from short bunches are considered. Alternatively, plasma focusing technology can be explored, as proposed in \cite{rosenzweig_demonstration_1990} and subsequently tested in multiple experimental facilities \cite{nakanishi_direct_1991,hairapetian_experimental_1994,govil_observation_1999}.

\begin{figure}
	\centering
	\includegraphics[width=0.45\textwidth]{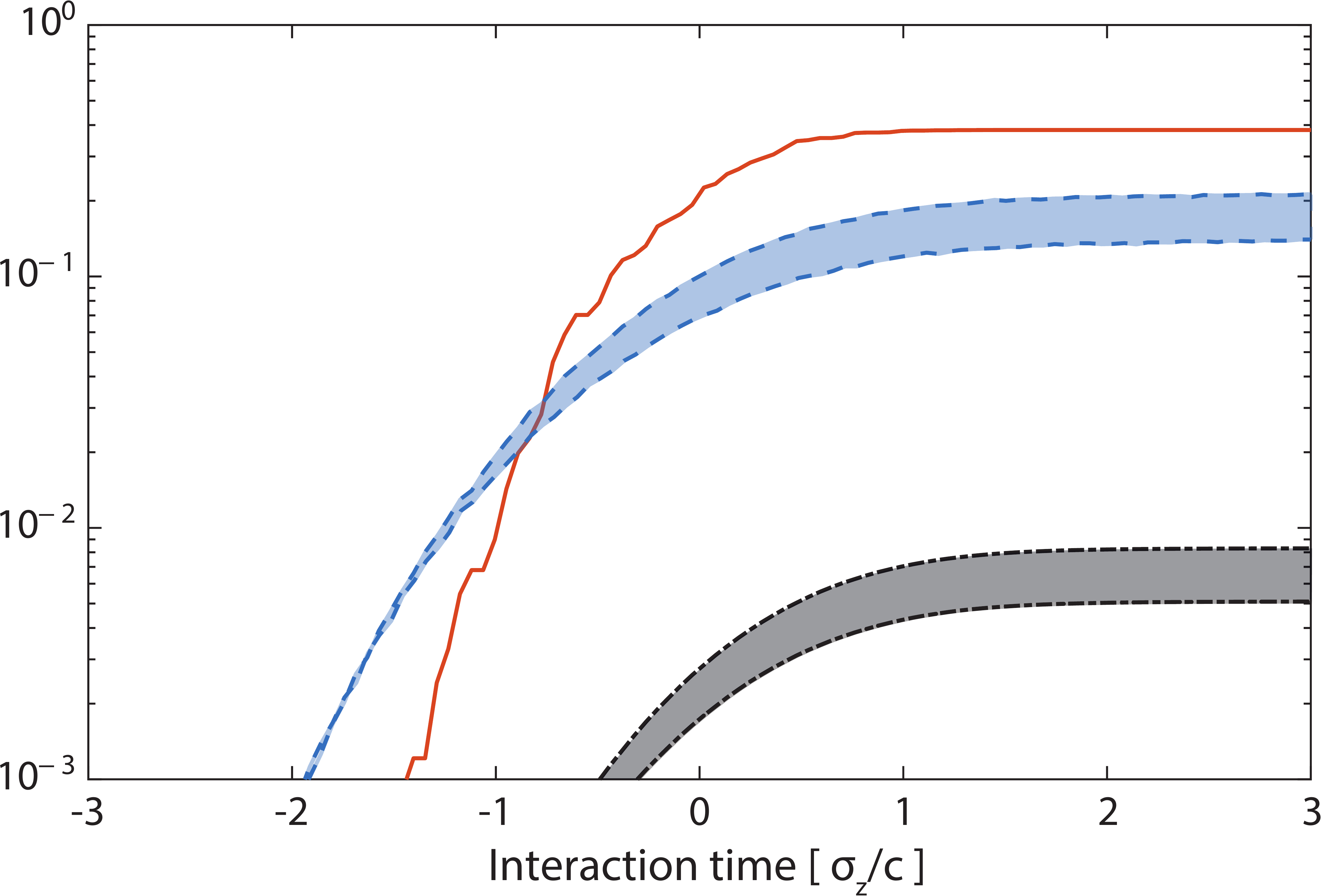} 
	\caption{\label{fig:3Dsim} Results of 3D simulations with OSIRIS-QED for the parameters of the NpQED collider in \tabref{tab:parameters}. $f_q$ (solid red): fraction of beam electrons experiencing $\alpha\chi^{2/3} \geq 1$; $f_\gamma$ (dashed blue): number of high-energy photons ($E_\gamma \geq 2 mc^2$) emitted per beam electron; $f_p$ (dotted black): number of secondary pairs per beam electron (all quantities are shown as a function of the beam crossing time). \hl{The shaded areas represent an estimate for the importance of nonperturbative quantum effects ($f_\gamma \sim 30\%$, $f_p \sim 25\%$). They were obtained by modifying the photon emission/pair production probabilities in OSIRIS-QED as explained in the main text and the Supplemental Material (upper curves: state-of-the-art simulation, lower curves: modified probabilities).}
	}
\end{figure}

\hl{Even though a complete engineering design of the accelerator layout requires further R\&D on the various subsystems -- including high brightness beam sources, advanced beam compression techniques, final focus and beam delivery system -- the NpQED collider parameters (\tabref{tab:parameters}) rely only on evolutionary improvements of existing technology.}

In order to confirm the possibility of reaching the regime $\alpha\chi^{2/3} \gtrsim 1$ with short, high current, colliding beams, we have performed 2D and 3D PIC simulations for the parameters of the NpQED collider in \tabref{tab:parameters}. We employed the massively parallel, fully relativistic and electromagnetic PIC code OSIRIS-QED \cite{grismayer_seeded_2017,grismayer_laser_2016,fonseca_osiris_2002}, which accounts self-consistently for the classical and the QED interaction between particles and fields (see Supplemental Material). \hl{The latter is taken into account by employing photon emission and pair production probabilities inside a constant field \cite{elkina_qed_2011,nerush_laser_2011,ridgers_dense_2012,gonoskov_extended_2015,lobet_generation_2017}. This so-called local constant field approximation (LCFA) is applicable here, as the formation length $l_f = c \Delta t \sim \gamma \lambdabar_{c} / \chi^{2/3} \sim \unit[1]{nm}$ \cite{baier_quantum_1989} is much smaller than the scale on which the field changes ($\sigma_z = \unit[10]{nm}$).} 

Figures\,\ref{fig:collider} and \ref{fig:3Dsim} illustrate the results of 3D simulations for electron beams with $\sigma_z = \unit[10]{nm}$ ($D= 10^{-3}$). The simulations confirm that these beam parameters provide a suitable configuration to probe the fully nonperturbative QED regime, as a large fraction of beam electrons (38\%) experience $\alpha\chi^{2/3} \geq 1$, while the beam energy losses are limited to $\lesssim 5\%$ (\figref{fig:2Dsim}). 

\begin{figure}
	\centering
	\vspace*{3pt}\includegraphics[width=0.45\textwidth]{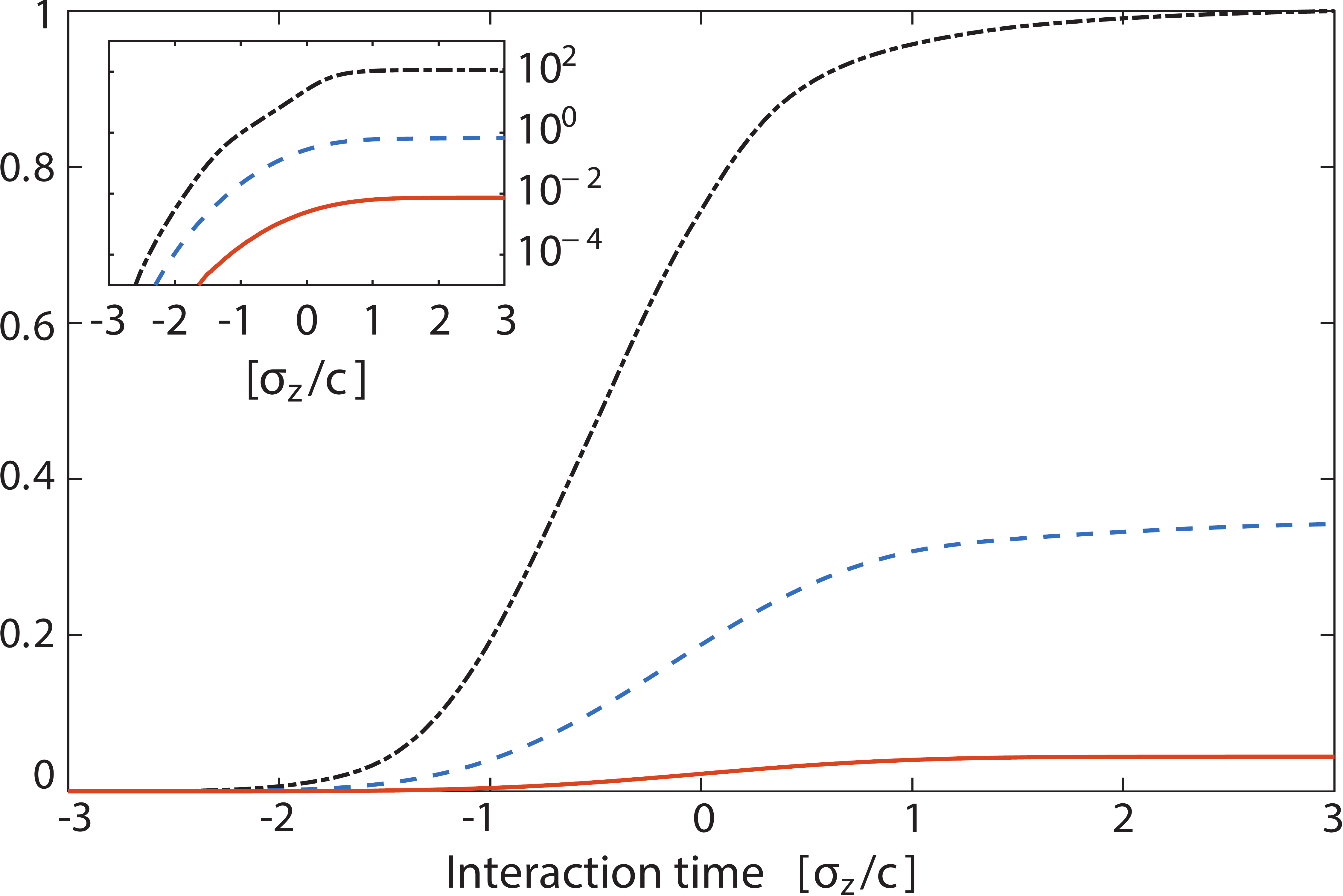}
 	\caption{\label{fig:2Dsim} Relative energy loss and number of secondary pairs $f_p$ (inset plot) for three beam lengths:  $\unit[10]{nm}$ (solid red, $D=10^{-3}$), $\unit[100]{nm}$ (dashed blue, $D=0.1$), and $\unit[500]{nm}$ (dotted back, $D=2.5$). In the latter case electron trajectories are severely modified, leading to strong disruption. The interaction results in a full-featured QED cascade, which multiplies the particle density and completely absorbs the energy of the initial beam.} 
\end{figure}

\hl{To estimate the importance of nonperturbative effects, we have phenomenologically taken into account that quantum corrections dynamically increase the effective electron/positron mass $m_*$ (see Supplemental Material)
\begin{gather}
\label{eqn:mstar}
m_*^2 = m^2 + \delta m^2, \quad \delta m^2 \approx 0.84 \alpha\chi^{2/3} m^2.
\end{gather}
Thus, we should replace $\chi$ by $\chi_* = \chi(m\to{}m_*)$ in the photon emission probability $W(\chi)$, i.e., employ $\widetilde{W}(\chi) = W[\chi_*(\chi)]$ instead. After an elementary calculation one obtains 
\begin{gather}
\label{eqn:chistarasym}
\chi^{\nfrac{2}{3}}_*(\chi) = \chi^{2/3}[1 + 0.84 \alpha \chi^{2/3}]^{-1}. 
\end{gather}
Note that corrections to the effective electron/positron mass are not the only consequence of the nonperturbative regime. However, a complete and rigorous nonperturbative calculation, e.g., by employing methods developed for strongly-coupled quantum field theories like QCD, is far beyond the scope of this Letter (e.g., truncated Schwinger-Dyson equations, resummation of certain diagram classes, renormalization-group techniques \cite{polonyi_lectures_2003,roberts_dyson-schwinger_1994,zinn-justin_perturbation_1981}). However, the above estimate allows us to anticipate the order of magnitude of nonperturbative corrections in the regime $\alpha\chi^{2/3} \lesssim 1$ (\figref{fig:3Dsim}), similar as a phenomenological recoil correction to the classical probabilities allows us to estimate the order of magnitude of quantum corrections in the regime $\chi \lesssim 1$ (see Supplemental Material).} 

The bunch length, and correspondingly the disruption parameter, significantly impact the dynamics of the beam-beam interaction. To quantify this, a series of 2D simulations for $\sigma_z = \unit[10 - 500]{nm}$ ($D = 10^{-3}$ -- $2.5$, \figref{fig:2Dsim}) has been performed. The results indicate that electron trajectories become considerably disrupted for $D \gtrsim 0.1$ and energy losses are no longer negligible ($>30\%$). Therefore, $D < 0.1$ is preferable, as it provides a clean experimental interaction for testing theoretical nonperturbative QED predictions. However, we note that the $D \gtrsim 1$ regime represents an interesting scientific frontier, where the interplay between collective and strong-field quantum processes determines the evolution of the system.

In summary, we have shown that the collision of tightly compressed and focused 100 GeV-class electron beams would offer a very promising configuration for probing the fully nonperturbative QED regime $\alpha\chi^{2/3} \gtrsim 1$. Until now, the physics above this threshold remains completely unexplored experimentally, and there is no theoretical framework to describe light-matter interaction at such extreme fields. Investigations of this qualitatively different regime, both theoretical and experimental, are bound to discover new physical phenomena and advance the understanding of nonperturbative physics at the field intensity frontier.

\begin{acknowledgements}
The authors would like to thank Antonino Di Piazza, Lance Dixon, Nathaniel J. Fisch, Michael Peskin, Tor Raubenhamer, and Ricardo A. Fonseca for useful discussions. This work was supported by: U.S. Department of Energy under contract number DE-AC02-76SF00515, U.S. DOE Early Career Research Program under FWP100331, Deutsche Forschungsgemeinschaft (DFG, German Research Foundation) ‚Äì 361969338 and 367991447, MEPhI Academic Excellence Project (Contract No. 02.a03.21.0005), Russian Fund for Basic Research (Grant 16-02-00963), Foundation for the Advancement of Theoretical Physics "BASIS" (Grant 17-12-276-1), European Research Council (ERC-2015-AdG Grant No. 695008), and FCT (Portugal) grants SFRH/IF/01780/2013 and PD/BD/114323/2016. Simulations were performed at the IST cluster (Lisbon, Portugal) and at MareNostrum (Spain) under a PRACE award.
\end{acknowledgements}

\end{document}

\title{On the Prospect of Studying Nonperturbative QED with Beam-Beam Collisions: Supplemental Material}

\author{V.~\surname{Yakimenko}}
\email{yakimenk@slac.stanford.edu}
\affiliation{SLAC National Accelerator Laboratory, Menlo Park, CA USA}
\author{S.~\surname{Meuren}}
\affiliation{Department of Astrophysical Sciences, Princeton University, Princeton, New Jersey USA}
\author{F.~Del Gaudio}
\affiliation{GoLP/Instituto de Plasmas e Fus\~ao Nuclear, Instituto Superior T\'ecnico, Universidade de Lisboa, Lisboa, Portugal}
\author{C.~\surname{Baumann}}
\affiliation{Heinrich-Heine-Universit\"at, D\"usseldorf, Germany}
\author{A.~\surname{Fedotov}}
\affiliation{National Research Nuclear University MEPhI, Moscow, Russia}
\author{F.~\surname{Fiuza}}
\affiliation{SLAC National Accelerator Laboratory, Menlo Park, CA USA}
\author{T.~\surname{Grismayer}}
\affiliation{GoLP/Instituto de Plasmas e Fus\~ao Nuclear, Instituto Superior T\'ecnico, Universidade de Lisboa, Lisboa, Portugal}
\author{M.\,J.~\surname{Hogan}}
\affiliation{SLAC National Accelerator Laboratory, Menlo Park, CA USA}
\author{A.~\surname{Pukhov}}
\affiliation{Heinrich-Heine-Universit\"at, D\"usseldorf, Germany}
\author{L.\,O.~\surname{Silva}}
\affiliation{GoLP/Instituto de Plasmas e Fus\~ao Nuclear, Instituto Superior T\'ecnico, Universidade de Lisboa, Lisboa, Portugal}
\author{G.~\surname{White}}
\affiliation{SLAC National Accelerator Laboratory, Menlo Park, CA USA}

\begin{abstract}
\end{abstract}

\date{\today}

\maketitle

\section{QED-PIC approach}

Two-dimensional and three-dimensional QED-PIC simulations were carried out using the fully relativistic and fully electromagnetic particle-in-cell (PIC) code OSIRIS 4.0 \cite{fonseca_osiris_2002}. The novel QED-PIC approach, which models both collective/plasma and quantum effects self-consistently, has been employed by several groups recently \cite{elkina_qed_2011,nerush_laser_2011,ridgers_dense_2012,gonoskov_extended_2015,lobet_generation_2017} and the implementation in OSIRIS follows a similar methodology (see also \cite{duclous_monte_2011,green_simla_2015,tamburini_laser-pulse-shape_2017}). OSIRIS-QED has been benchmarked with other existing QED-PIC codes and used to study and model QED cascades \cite{grismayer_laser_2016,grismayer_seeded_2017} and beam-beam collisions in the weak quantum regime ($\chi\sim 1$). The results presented here were also reproduced by the code VLPL \cite{baumann_influence_2016}.

QED-PIC simulations take advantage of the fact that in strong electromagnetic background fields the classical dynamics of electrons, positrons and the background field itself is much slower than the QED processes, which change the number of particles. Therefore, it is possible to model the latter with a Monte-Carlo event generator. At every time step of the PIC loop, which solves Maxwell’s equations and the Lorentz force self-consistently, the probability for photon emission by electrons/positrons and pair production by high-energy photons is calculated. Depending on the parameter regime and the context, field-induced photon emission is also called (nonlinear/multiphoton) Thomson/Compton scattering, (quantum) synchrotron radiation or beamstrahlung. Furthermore, field-induced pair production is also called coherent pair production or referred to as the (nonlinear/multiphoton) Breit-Wheeler process. If a QED transition takes place, the energy of the created particles is determined randomly from the respective QED distributions. For highly energetic particles the energy completely determines the kinematics, as the propagation direction approximately coincides with the one of the incoming particle. 
Two of the most used codes in the high energy physics community to model beam-beam and laser-beam interactions in the QED regime are CAIN \cite{CAIN} and GUINEA-PIG \cite{schulte_diss_1996}. However, the approximations employed in these codes are only valid in the low disruption regime, where simulations performed with OSIRIS-QED and GUINEA PIG are in good agreement. The benefit of using OSIRIS-QED simulations lies in the possibility of accessing higher disruption regimes and to model the self-consistent modification of the beam fields due to the emission of secondary electron-positron pairs. 

PIC simulations with long beam simulations require considerable computational efforts, rendering the full 3D modelling unpractical. It is, however, possible to investigate the impact of longer beams and larger disruption parameters on beam-beam collisions by leveraging on 2D QED-PIC simulations. All simulations are performed for 125 GeV beams, with a spot size nm, and a beam current of 1.7 MA for the 3D case. The beam current in 2D is set to match the peak value of the 3D field. The simulation parameters are summarized in Table~\ref{tab:param}.

\begin{table}
\caption{\label{tab:param}\textbf{Summary of the simulation parameters.} Here, $\sigma_z$ denotes the length of the beam, $L_x$, $L_y$, $L_z$ are the physical length, width and height of the simulation box, $N_x$, $N_y$, $N_z$ represent the number of grid cells used in each direction, $dt$ is the time step measured in units of the inverse of the beam plasma frequency $\omega_b$ and \#PPC the number of particles used per cell.}
\centering
\begin{tabular}{cccccc}
\toprule
\rotatebox{90}{Dim} & \minibox[c]{$\sigma_z$\\ $\mathrm{[nm]}$} & \minibox[c]{$L_x/L_y/L_z$\\ $\mathrm{[c/\omega_b]}$} & \minibox[c]{$N_x/N_y/N_z$\\$\phantom{[]}$} & \minibox[c]{$dt$\\ $\mathrm{[\omega_b^{-1}]}$} & \rotatebox{90}{\hspace*{-5pt}\#PPC} \\
\midrule
3D & 10 & 180/100/100 & 900/500/500 & 0.114 & 1 \\
2D & 10 & 140/40/n.a. & 1400/800/n.a. & 0.069 & 16 \\
2D & 100 & 1220/80/n.a. & 6100/400/n.a. & 0.13 & 16\\
2D & 500 & 6020/200/n.a. & 12040/800/n.a. & 0.219 & 1\\
\bottomrule
\end{tabular}
\end{table}

For the QED-PIC approach to be valid, the applicability of the local constant field approximation (LCFA) for the field-induced QED processes is crucial \cite{ritus_1985,elkina_qed_2011}. In the quantum regime $\chi \gtrsim 1$ the formation length $l_f$ for photon emission and pair production scales as $l_f \sim \gamma \lambdabar_{c} / \chi^{2/3}$, where $\lambdabar_c \approx \unit[3.8 \times 10^{-13}]{m}$ denotes the reduced Compton length and $\gamma = \eps / (mc^2)$ the gamma factor of an electron/positron with energy $\eps$ \cite{baier_quantum_1989}. As the collective background field changes on the scale $\sigma_z$, the condition $l_f \ll \sigma_z$ should be satisfied. For $\eps \approx \unit[125]{GeV}$ and $\chi \sim 10^3$ this is the case, as $l_f \sim \unit[1]{nm} \ll \unit[10]{nm} \lesssim \sigma_z$. 

In addition to field-induced processes, also binary collisions between individual electrons, positrons, and photons can, in principle, occur. As the cross section for such processes is at most of order $r_e^2$, the corresponding mean free path is given by $\lambda_{\mathrm{re}} \sim 1/(nr_e^2)$, where $n$ denotes the beam particle density and $r_e = \alpha \lambdabar_c \approx \unit[2.8 \times 10^{-15}]{m}$ the classical electron radius \cite{landau_quantum_1981}. For the parameters considered, $n \approx \unit[10^{26}]{cm^{-3}}$ and $\alpha\chi^{2/3} \sim 1$, implying $\lambda_{\mathrm{bf}} \sim \alpha^2 \lambda_{\mathrm{re}}$, where $\lambda_{\mathrm{bf}} \sim l_f / \alpha$ denotes the mean free path for field-induced QED processes. Therefore, binary particle-particle interactions are negligible here.

\section{Quantum corrections in strong fields}

In vacuum the interaction between light and matter is always ``perturbative'', as the fine-structure constant $\alpha = \nfrac{e^2}{(4\pi)} \approx 1/137$ is small (Heaviside and natural units are used in this supplement, i.e., $\epsilon_0 = \hbar = c = 1$; $e>0$ is the elementary charge). Therefore, quantum fluctuations only induce small corrections, which result in a running coupling constant $\alpha(\mu_{\text{ms}}^2)$ with a logarithmic dependence on the energy scale $\mu_{\text{ms}}^2$ \cite{landau_quantum_1981}. Correspondingly, QED radiative corrections are well understood in the low-intensity limit, where QED facilitates extremely precise theoretical predictions.

The situation changes profoundly in the presence of a strong electromagnetic background field, described by the (classical) field tensor $F^{\mu\nu} = \del^\mu A^\nu - \del^\nu A^\mu$. A constant and uniform field provides two Lorentz and gauge invariant dimensionless quantities (see, e.g., \cite{landau_quantum_1981}, \S\,101)
\begin{gather}
\label{eqn:fieldinvariants}
f = \frac{1}{F^2_{\text{cr}}} F_{\mu\nu} F^{\mu\nu}, \quad g = \frac{1}{F^2_{\text{cr}}} \eps^{\mu\nu\rho\sigma} F_{\mu\nu} F_{\rho\sigma}.
\end{gather}
Here, $\eps^{\mu\nu\rho\sigma}$ denotes the Levi-Civita totally antisymmetric tensor, $F_{\text{cr}} = \nfrac{m^2}{e}$ the QED critical field, and $m$ the electron/positron mass. Assuming that in addition to the field a particle or photon with four-momentum $p^\mu = (\eps,\spvec{p})$ is present, a third invariant -- the so-called quantum or beamstrahlung parameter -- can be constructed (see, e.g., \cite{landau_quantum_1981}, \S\,101)
\begin{gather}
\label{eqn:chi}
\chi = \frac{\sqrt{-(F^{\mu\nu}p_{\nu})^2}}{F_{\text{cr}} m} \sim \frac{\eps}{m} \frac{F}{F_{\text{cr}}}
\end{gather}
(sometimes an index is added to clarify which four-momentum is used to construct $\chi = \chi_p$). The order of magnitude estimation in \eqref{eqn:chi} supposes that the particle or photon is highly energetic ($\eps \gg m$) and $F \sim \abs{E}$ or $F \sim \abs{B}$, depending on which field is larger in the laboratory frame. 

Realistic laboratory fields are much smaller than the QED critical field $F_{\text{cr}}$, implying $f,g \sim (\nfrac{F}{F_{\text{cr}}})^2 \ll 1$ [see \eqref{eqn:fieldinvariants}]. However, they can become comparable with $F_{\text{cr}}$ in a boosted frame, where they are enhanced by the Lorentz gamma factor $\eps/m$ [see \eqref{eqn:chi}], i.e., $\chi \gtrsim 1$. Therefore, we focus here on the situation $\chi \gtrsim 1$, $1 \gg f,g$, where $\chi$ is the only relevant field-related parameter (see, e.g., \cite{landau_quantum_1981}, \S\,101 and \cite{ritus_1985}). Hence, it is sufficient to consider a constant-crossed field ($f=g=0$) in the following.

According to the Ritus-Narozhny conjecture \cite{ritus_radiative_1972,narozhny_expansion_1980,fedotov_conjecture_2017} the true expansion parameter of QED in the limit $\chi \gg 1$ is not $\alpha$, as in vacuum, but $\alpha \chi^{2/3}$. Correspondingly, we expect that the predictions of QED change qualitatively in the ``fully non-perturbative regime'' $\alpha \chi^{2/3} \gtrsim 1$. 

In this supplement we briefly summarize how the parameter $\alpha \chi^{2/3}$ emerges and provide a back-of-the-envelope intuitive derivation from fundamental principles. For more details we refer the reader to review articles and textbooks, e.g., \cite{ritus_1985,fradkin_quantum_1991,di_piazza_extremely_2012}.

\subsection{Photon wave equation}

In quantum mechanics the fundamental properties of all particles are determined by wave equations, which are, in principle, modified by quantum fluctuations. In vacuum, however, their influence vanishes after the theory is properly renormalized. This changes in the presence of a strong background field, which polarizes the quantum vacuum. Correspondingly, the complete wave equation for the ``photon wave function'' $\Phi^\mu(x)$ is given by (see, e.g., \cite{landau_quantum_1981}, \S\,103 and \cite{bialynicki-birula_photon_1996})
\begin{gather}
\label{eqn:photonwaveeq}
-\del^2 \Phi^\mu(x) = \int d^4y \, \mc{P}^{\mu\nu}(x,y) \Phi_{\nu}(y), 
\end{gather}
where $\mc{P}^{\mu\nu}(x,y)$ is the renormalized polarization operator in position space (see Fig.\,2 of the main text). 

In classical electrodynamics the photon wave function $\Phi^\mu(x)$ corresponds to the four-potential of the radiation field. In Lorentz gauge $\del_\mu \Phi^\mu(x) = 0$ it obeys the inhomogeneous wave equation $\del^2 \Phi^\mu(x) = J^\mu$, where $J^\mu$ represents the four-current density. As the current on the right-hand-side of \eqref{eqn:photonwaveeq} is induced by the photon field itself via $\mc{P}^{\mu\nu}(x,y)$, the strong-field vacuum resembles a polarizable medium. This profound analogy between a photon propagating through a strong-field vacuum and an electromagnetic wave inside a polarizable medium is further explored in \secref{sec:plasmafrequency}. 

\subsection{Fermion wave equation}

Similar as for photons, quantum fluctuation also modify the wave equation for electrons/positrons. Inside a background field the so-called Dirac-Schwinger equation has to be solved (see, e.g., \cite{landau_quantum_1981}, \S\,109 and \cite{schwinger_gauge_1951})
\begin{gather}
\label{eqn:diracschwingereq}
\big[ i\s{\del} + e\s{A}(x) - m \big] \Psi(x) = \int d^4y \, \mc{M}(x,y) \Psi(y). 
\end{gather}
Here, $\Psi(x)$ denotes the electron/positron Dirac spinor field, $A^\mu(x)$ the classical four-potential of the background field, and $\mc{M}(x,y)$ the renormalized mass operator in position space (see Fig.\,2 of the main text). Furthermore, the Feynman slash notation is employed, i.e., $\s{\del} = \gamma^\mu \del_\mu$ and $\s{A} = \gamma^\mu A_\mu$, where $\gamma^\mu$ are the Dirac matrices.

If quantum corrections are dropped [$\mc{M}(x,y) \equiv 0$], \eqref{eqn:diracschwingereq} reduces to the ordinary Dirac equation with a classical background field. Solutions of \eqref{eqn:diracschwingereq} with $\mc{M}(x,y)\equiv 0$ but which take $A^\mu(x)$ into account exactly are called Furry-picture \cite{furry51} or dressed states and are denoted by double lines in Feynman diagrams.

\subsection{Breakdown of perturbation theory}

Besides changing the wave equations for electrons/positrons and photons, quantum fluctuations also modify their interaction vertex. In principle, we have to solve so-called Dyson-Schwinger equations self-con\-sis\-tent\-ly in order to find the fundamental eigenstates of the theory (see, e.g., \cite{landau_quantum_1981}, \S\,107). In practice, an exact solution is impossible and a certain truncation scheme has to be applied.

In leading-order perturbation theory (leading-order loop expansion) the mass operator contains only vacuum photons [i.e., \eqref{eqn:photonwaveeq} with $\mc{P}^{\mu\nu}(x,y) = 0$] and the polarization operator only field-dressed electrons/positrons [i.e., \eqref{eqn:diracschwingereq} with $\mc{M}(x,y) = 0$]. Furthermore, the vertex correction is negligible for both. Therefore, the wave equations for photons and electrons/positrons decouple and can be solved independently. 

According to the Ritus-Narozhny conjecture \cite{ritus_radiative_1972,narozhny_expansion_1980,fedotov_conjecture_2017}, this perturbative analysis breaks down in the regime $\alpha \chi^{2/3} \gtrsim 1$, allowing the predictions of the theory to change qualitatively. In the next sections we will first summarize the well established perturbative results for photons and electrons/positrons inside a constant-crossed field and then present arguments which substantiate the Ritus-Narozhny conjecture. 

\subsection{Comparison with supercritical magnetic fields}

A similar breakdown of perturbation theory is predicted for electrons/positrons occupying the lowest Landau level of a supercritical magnetic field [$B \gg B_{cr} = m^2 c^2/({e \hbar}) \approx 4.41 \times 10^9 {T}$]. In this case, too, the appearance of non-perturbative effects coincides with a change in hierarchy between the electron/positron rest mass m and the field-induced quantum corrections \cite{gusynin_dynamical_1999,kuznetsov_electron_2002}. The latter, however, scale only logarithmically with the field ($\log⁡(B / B_cr)$), \cite{jancovici_radiative_1969}. Therefore, even the strongest magnetic fields in the universe do not suffice to induce substantial corrections \cite{cerutti_electrodynamics_2017}. The case considered here is complementary and qualitatively different, as it corresponds on the contrary to electrons/positrons occupying very high Landau levels, where a power-law scaling ($\chi^2/3$) is encountered instead. In fact, this is the reason why the collider setup considered here is capable for probing the fully non-perturbative regime of QED. Furthermore, only electrons/ positrons which are not in the ground state can emit photons and produce pairs and thus provide two accessible observables which are affected by radiative corrections. 

\section{Field-induced effective mass}

\subsection{Photons}

On the scale on which a field can be considered constant and homogeneous, the photon four-momentum is conserved and we can solve the wave equation [see \eqref{eqn:photonwaveeq}] with a plane-wave ansatz in momentum space [$\Phi^\mu(x)  = \eps_j^\mu e^{-iQ_j x}$]. Here, $Q_j^\mu = (\omega_{j},\spvec{Q}_j)$ denotes the effective four-momentum of the photon inside the background field ($Q_j^\mu \to q^\mu$ with $q^2=0$ if $F^{\mu\nu}\to 0$; see \cite{bragin_high-energy_2017} and the references therein). 

The two (pseudo) four-vectors $\eps_1^\mu \sim F^{\mu\nu} q_{\nu}$ and $\eps_2^\mu \sim \eps^{\mu\nu\rho\sigma} q_{\nu} F_{\rho\sigma}$ represent the polarization four-vectors of the two transverse eigenmodes of the photon field. The dispersion relations $\omega^2_{j} = m^2_{\gamma,j} + \spvec{Q}_j^2$ of the two eigenmodes differ, i.e., the effective photon mass $m_{\gamma,j}$ depends on the polarization state [see, e.g., \cite{ritus_1985}, page 601, Eq.\,(87) and \cite{narozhnyi_propagation_1968}]
\begin{gather}
\label{eqn:ccfphotonmasses}
\begin{bmatrix}m^2_{\gamma,1}\\m^2_{\gamma,2}\end{bmatrix}
=
\alpha\, \frac{m^2}{3\pi} \int_{-1}^{+1} dv \, \begin{bmatrix}(w-1)\\(w+2)\end{bmatrix} \Big(\frac{\chi}{w}\Big)^{2/3} f'(\rho),
\end{gather}
where $\nfrac{1}{w} = (\nfrac{1}{4})(1-v^2)$, $\rho = \big(\nfrac{w}{\chi}\big)^{2/3}$, $f(\rho) = \pi \Gi(\rho) + i\pi \Ai(\rho)$ is the Ritus function, and $\Ai$ and $\Gi$ denote the Airy and the Scorer function, respectively \cite{olver_nist_2010} (the derivative of a function is indicated by a prime, $\chi = \chi_{q}$).

Inside a background field the photon can decay into an electron-positron pair. Correspondingly, the effective photon mass has an imaginary part, which corresponds to the total photon decay probability \cite{ritus_1985}. 

Furthermore, the polarization dependence of the photon implies that the strong-field vacuum exhibits birefringence and dichroism \cite{narozhnyi_propagation_1968}. These effects are experimentally observable consequences of field-induced radiative corrections to the photon wave function (see, e.g., \cite{bragin_high-energy_2017} and the references therein). 

From \eqref{eqn:ccfphotonmasses} we obtain the following asymptotic expression for the photon masses ($\chi \gg 1$) 
\begin{gather}
\label{eqn:ccfphotonmasses_chilarge}
\begin{bmatrix}m^2_{\gamma,1}\\m^2_{\gamma,2}\end{bmatrix}
\approx
\alpha \chi^{2/3} m^2 \, \frac{3^{7/6}\Gamma^4(2/3)}{14\pi^2} \begin{bmatrix}2\\3\end{bmatrix} (1 - i\sqrt{3}),
\end{gather}
where $\Gamma(z)$ denotes the Gamma function \cite{olver_nist_2010}.

As discussed in the main text, the quantum scale $\mu^2 = \alpha \chi^{2/3} m^2$ is independent of $m$ and thus purely background-field induced. As it depends on $\hbar$, it represents a genuine quantum effect. For $\alpha \chi^{2/3} \gtrsim 1$ the photon mass generated by the classical background field via quantum corrections is of the same order as the electron/positron vacuum mass. Therefore, it is to be expected that in this regime the interaction between light and matter changes profoundly [note that the results given in Eqs.\,(\ref{eqn:ccfphotonmasses}) and (\ref{eqn:ccfphotonmasses_chilarge}) are only rigorously valid in the regime $\alpha \chi^{2/3} \ll 1$].

\subsection{Fermions}

The analysis for electrons/positrons is very similar to the one for photons summarized in the previous section. Inside the background field electrons/positrons obtain an effective four-momentum $P_\pm^\mu = (\eps_\pm,\spvec{P}_\pm)$, which reduces to $p^\mu$ ($p^2 = m^2$) in the absence of the field ($P^\mu_\pm \to p^\mu$ if $F^{\mu\nu}\to 0$; see \cite{meuren_quantum_2011} and the references therein). The spin eigenstates are defined by the canonical quantization axis $s^\mu \sim \eps^{\mu\nu\rho\sigma} p_{\nu} F_{\rho\sigma}$. Like for photons, the dispersion relations $\eps^2_{\pm} = m^2_{\pm} + \spvec{P}^2_\pm$ for electrons/positrons with spin up and spin down relative to the canonical quantization axis differ, i.e., the effective electron/positron mass $m_\pm$ is spin dependent [see, e.g., \cite{ritus_1985}, page 595, Eq.\,(54) and \cite{meuren_quantum_2011}]
\begin{multline}
\label{eqn:ccfelectronmasses}
m_\pm^2 =  m^2 + \frac{\alpha m^2}{\pi}  \, \int_0^\infty \frac{du}{(1+u)^3} \bigg[ \pm  \chi z f(z) \\+ \frac{5+7u+5u^2}{3} \lb\frac{\chi}{u}\rb^{2/3} f'(z) \bigg],
\end{multline}
where $z= (\nfrac{u}{\chi})^{2/3}$. For $\chi = \chi_{p} \gg 1$ we find the same scaling for the induced quantum corrections as for photons 
\begin{multline}
\label{eqn:ccfelectronmasses_chilarge}
m_\pm^2 - m^2 \approx \alpha \chi^{2/3} m^2 \, \frac{14}{9} \frac{\Gamma(2/3)}{3^{1/3} \sqrt{3}} (1-i\sqrt{3})
\\\pm \alpha \chi^{1/3} m^2 \, \frac{1}{9} \frac{\Gamma(1/3)}{3^{2/3} \sqrt{3}} (1+i\sqrt{3})
\end{multline}
[note that the results given in Eqs.\,(\ref{eqn:ccfelectronmasses}) and (\ref{eqn:ccfelectronmasses_chilarge}) are only rigorously valid if $\alpha \chi^{2/3} \ll 1$].

For electrons/positrons the imaginary part of the effective mass corresponds to the total radiation probability \cite{ritus_1985} and the mass difference between the two eigenmodes gives rise to phenomena analogous to vacuum birefringence and dichroism \cite{meuren_quantum_2011}. 

In the regime $\alpha \chi^{2/3} \gtrsim 1$ the quantum mass scale $\mu^2 = \alpha \chi^{2/3} m^2$ starts to dominate the effective electron masses $m_\pm$. Correspondingly, the properties of electrons/positrons and photons are now determined primarily by quantum corrections and start to deviate significantly from those well known and studied in ordinary QED.

\section{Estimate of nonperturbative effects}

Even though an ab initio nonperturbative calculation of the photon emission/pair production probability in the regime $\alpha\chi^{2/3} \gtrsim 1$ is far beyond the scope of the present work, we will now phenomenologically estimate the order of magnitude of nonperturbative corrections in the ``threshold regime'' $\alpha\chi^{2/3} \lesssim 1$. To this end we note that asymptotically (i.e., for $\chi \gg 1$) perturbation theory predicts [see \eqref{eqn:ccfphotonmasses_chilarge} and \eqref{eqn:ccfelectronmasses_chilarge}]
\begin{gather}
\label{eqn:asymptoticeffectivemasses}
m^2_\gamma \approx 0.2 \alpha \chi^{2/3} m^2,
\quad
m^2_* \approx  m^2 + 0.84 \alpha \chi^{2/3} m^2,
\end{gather}
where $m_\gamma$ ($m_*$) denotes the real part of the polarization (spin) averaged effective photon (electron/positron) mass. 

As the dynamically generated photon mass is much smaller than the correction to the electron/positron mass, we assume in the following that the latter represents the dominant effect. Therefore, it is to be expected that in the regime $\alpha \chi^{2/3} \gtrsim 1$ the QED critical field should be defined with respect to the effective electron/positron mass $m^*$ instead of $m$. 

As the field-free mass scale $m$ is irrelevant for quantum fluctuations in the regime $\alpha \chi^{2/3} \gtrsim 1$, it is natural to assume that nonperturbative effects change \eqref{eqn:asymptoticeffectivemasses} in a way such that
\begin{gather}
\label{eqn:asymptoticeffectivemasse}
m^2_* =  m^2 + \delta m^2, \quad \delta m^2 \approx 0.84 \alpha \chi_*^{2/3} m_*^2
\end{gather}
represents a good estimate, at least for $\alpha \chi^{2/3} \lesssim 1$. Here, %[$\chi_* = \chi(m\to{}m_*)$]. Finally, we obtain
\begin{gather}
\label{eqn:chistarasym}
\chi^{\nfrac{2}{3}}_* = \chi^{\nfrac{2}{3}} \frac{m^2}{m_*^2}  \approx \frac{\chi^{2/3}}{1 + 0.84 \alpha \chi^{2/3}}. 
\end{gather}
We point out that the asymptotic behaviour of \eqref{eqn:chistarasym} for $\chi \to \infty$ should not be taken too seriously, as we only expect this result to be indicative for $\alpha \chi^{2/3} \lesssim 1$.

\begin{figure}
	\centering
	\includegraphics[scale=1]{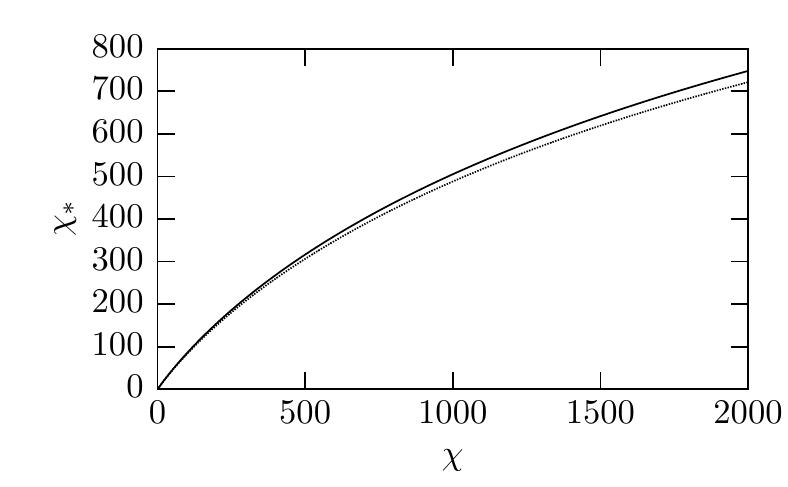}
	\caption{\label{fig:effectivechi}Effective electron/positron quantum parameter $\chi_*$ as a function of the perturbative value $\chi$. Solid curve: self-consistent numerical solution for $\chi_*$ based on \eqref{eqn:ccfelectronmasses}. Dotted curve: analytic approximation given in \eqref{eqn:chistarasym}. As the effective electron/positron mass $m_*$ increases with $\chi$, the effective QED critical field becomes larger and thus $\chi_* < \chi$. As a consequence, we expect that the photon emission and the pair production probabilities are attenuated.}
\end{figure}

As shown in \figref{fig:effectivechi}, the analytic result given in \eqref{eqn:chistarasym} is in very good agreement with a self-consistent numerical calculation based on the full leading-order expression for the (spin averaged) effective electron-positron mass given in \eqref{eqn:ccfelectronmasses}.

\subsection{Photon emission}

Based on the above considerations, $\chi_*$ represents the effective value of the quantum parameter which is experienced by an electron/positron. Correspondingly, it is reasonable to assume that the photon emission probability is now given by $\widetilde{W}(\chi) = W[\chi_*(\chi)]$, where $W(\chi)$ represents the corresponding perturbative result and $\chi_*(\chi)$ is given in \eqref{eqn:chistarasym}.

Inside a constant field one obtains the following radiation intensity spectrum [see, e.g., \cite{ritus_1985}, page 559, Eq.\,(52)]
\begin{multline}
\label{eqn:nlcscc_intensityspectrum_domega}
\frac{dI_{\text{rad}}}{d\omega dt}(\chi,\omega)  = - \alpha \frac{m^2}{\eps} \frac{u}{1+u} \Bigg\{ \int_z^\infty dv \, \Ai(v) \\+ \frac{\Ai'(z)}{z} \lsb 2 + \frac{u^2}{(1+u)} \rsb \Bigg\},
\end{multline}
where $\omega$ and $\eps$ denote the energy of the emitted photon and the incoming electron, respectively, $u = \omega/(\eps-\omega)$ and $z = (\nfrac{u}{\chi})^{\nfrac{2}{3}}$. The total emitted power (by a single electron/positron) is thus given by
\begin{gather}
\label{eqn:nlcscc_totalintensity}
P(\chi) = \frac{dI_{\text{rad}}}{dt} = \int_0^\infty d\omega \,  \frac{dI_{\text{rad}}}{d\omega dt}(\chi,\omega).
\end{gather}

\begin{figure}
	\centering
	\includegraphics[scale=1]{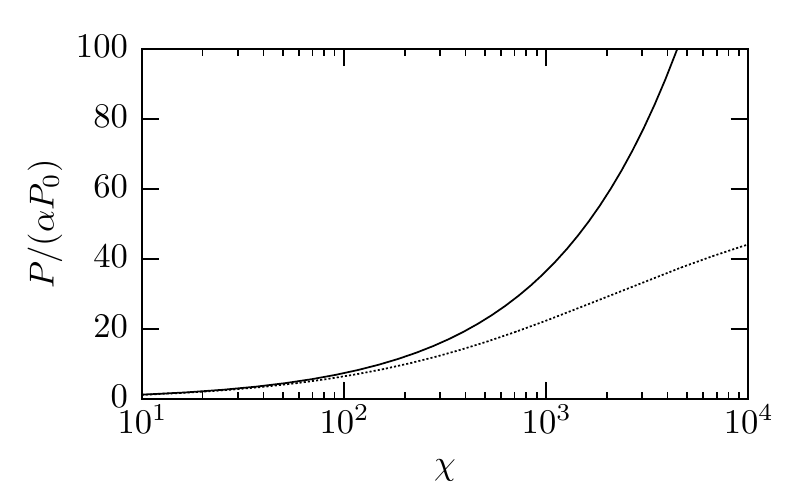}
	\caption{\label{fig:totalemission}Total radiation power as a function of $\chi$ [$P_0 = m^2 = (mc^2)^2/\hbar \approx \unit[6.4 \times 10^7]{W}$]. Solid curve: quantum result [see \eqref{eqn:nlcscc_totalintensity}], which neglects the increase of the electron/positron mass. Dotted curve: quantum result [see \eqref{eqn:nlcscc_totalintensity}] with the replacement $\chi\to\chi_*$, with $\chi_*$ given by \eqref{eqn:chistarasym}. Correspondingly, we expect a measurable suppression of the total energy loss for $\chi \gtrsim 10^3$.}
\end{figure}

In \figref{fig:totalemission} this perturbative result for $P(\chi)$ is compared with the expected nonperturbative correction $\widetilde{P}(\chi) = P[\chi_*(\chi)]$. Thus, we expect that nonperturbative effects attenuate radiative energy losses.

\subsection{Pair production}

The extension of the above consideration to pair production is straight forward. The most important modification arises from the fact that the created electron-positron pair has less energy than the decaying photon. For simplicity and definiteness, we assume that $\chi = \chi_\gamma /2$, where $\chi_\gamma$ and $\chi$ denote the quantum parameter for the incoming gamma photon and the created electron/positron, respectively. Finally, we obtain
\begin{gather}
\label{eqn:chigammastar}
\chi^{2/3}_{\gamma*} = \chi^{2/3}_{\gamma} \frac{m^2}{m_*^2} = \frac{\chi^{2/3}_{\gamma}}{1 + 0.84 \alpha (\chi_{\gamma}/2)^{2/3}}.
\end{gather}
To estimate nonperturbative modifications to the perturbative pair production probability $W_{\pm}(\chi)$ we replace it with $\widetilde{W}_{\pm}(\chi) = W_{\pm}[\chi_{\gamma*}(\chi_\gamma)]$.

%\subsection{Plausibility of the estimates}
\subsection{Comparison with the classical to quantum transition}

To substantiate that the above heuristic treatment of nonperturbative quantum effects results in meanigful estimates, we consider in the following the transition from classical ($\chi\ll 1$) to quantum ($\chi \gtrsim 0.1$) synchrotron radiation. 

Classically, the photon recoil is negligible and instead of \eqref{eqn:nlcscc_intensityspectrum_domega} we obtain [see, e.g., \cite{baier_electromagnetic_1994}, page 25, Eq.\,(1.92) and Eq.\,(1.94)]
\begin{gather}
\label{eqn:nlcscc_intensityspectrum_classical}
\frac{dI^{\text{class}}_{\text{rad}}}{d\omega dt}(\chi,\omega)  = - \alpha \frac{m^2}{\eps} u \bigg[ \int_z^\infty dv \, \Ai(v) + 2 \frac{\Ai'(z)}{z} \bigg],
\end{gather}
where $u = \nfrac{\omega}{\eps}$. Correspondingly,
\begin{gather}
\label{eqn:nlcscc_totalintensity_classical}
P(\chi) = \frac{dI^{\text{class}}_{\text{rad}}}{dt} = \int_0^\infty d\omega \,  \frac{dI^{\text{class}}_{\text{rad}}}{d\omega dt}(\chi,\omega) = \frac{2}{3} \alpha P_0 \chi^2,
\end{gather}
where $P_0 = m^2 = (mc^2)^2/\hbar \approx \unit[6.4 \times 10^7]{W}$. 

\begin{figure}
	\centering
	\includegraphics[scale=1]{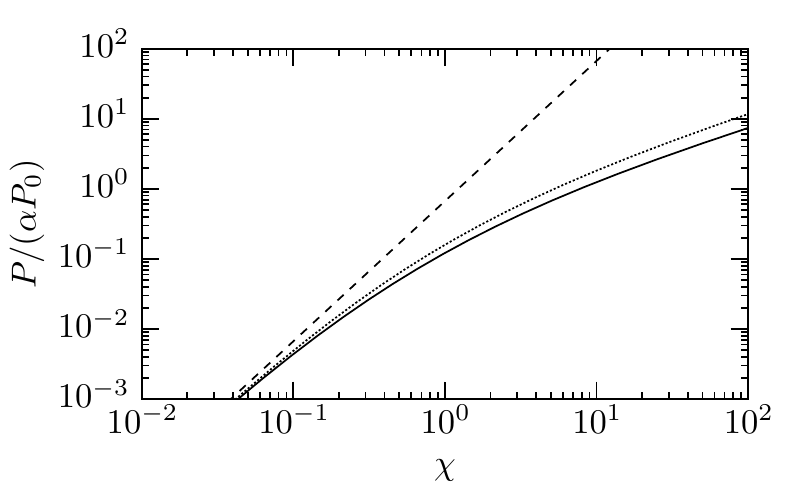}
	\caption{\label{fig:totalemissionclassical}Total radiation power as a function of $\chi$. Dashed curve: classical result [\eqref{eqn:nlcscc_totalintensity_classical} with $u=\omega/\eps$]. Dotted curve: classical result with phenomenological recoil correction [\eqref{eqn:nlcscc_totalintensity_classical} with $u = \omega/(\eps-\omega)$]. Solid curve: full quantum result [see \eqref{eqn:nlcscc_totalintensity}].}
\end{figure}

In order to phenomenologically estimate the importance of quantum corrections, i.e., the difference between \eqref{eqn:nlcscc_totalintensity} and \eqref{eqn:nlcscc_totalintensity_classical}, we start from \eqref{eqn:nlcscc_intensityspectrum_classical}, which is a result of classical electrodynamics. Assuming that the most important quantum effect is the recoil of the emitted photon, it is natural to apply the replacement $\eps \to \eps - \omega$ in the (classical) definition of $u = \omega/\eps$. As an important consequence energy conservation is ensured ($u$ diverges in the limit $\omega\to\eps$, which exponentially suppresses the radiation probability). 

By comparing \eqref{eqn:nlcscc_intensityspectrum_domega} with \eqref{eqn:nlcscc_intensityspectrum_classical} we see that quantum corrections also modify the preexponential structure of the emission probability
\begin{gather}
u \to u(1+u)^{-1}, 
\quad
2 \to 2 + u^2(1+u)^{-1}.
\end{gather}
In the regime $\chi \lesssim 1$, however, the replacement $u = \omega/\eps \to u = \omega/(\eps-\omega)$ is the most important change, as it leads to an exponential suppression of the radiation probability in the limit $\omega \to \eps$. As shown in \figref{fig:totalemissionclassical}, the recoil-corrected classical result agrees qualitatively with the full quantum calculation and represents even a very good quantitative approximation in the regime $\chi \lesssim 1$. Note that it was obtained directly from the classical result, i.e., without carrying out a full quantum calculation.

\section{Classical vs. quantum mass dressing}
\label{sec:classicalmassdressing}

So far, we reviewed the field-induced quantum corrections to the masses of electrons/positrons and photons, which are of the order $\mu^2 = \alpha \chi^{2/3} m^2$ if $\chi \gg 1$. It is instructive to compare this ``quantum mass dressing'' with the famous classical ``dressed mass'' or quiver energy of an electron/positron inside a plane-wave background field with characteristic frequency $\omega_k$ \cite{brown_interaction_1964}. It turns out that both effects have a similar physical origin.

For definiteness, we consider a linearly polarized plane wave background field, i.e., a field described by the four potential $A^\mu(\phi) = a^\mu \psi(\phi)$, where $\phi = kx$ denotes the phase of the plane-wave field [$k^\mu = (\omega_k, \spvec{k})$], $f^{\mu\nu} = k^\mu a^\nu - k^\nu a^\mu$ determines the magnitude of the corresponding field tensor, and $\psi(\phi) \sim 1$ is a shape function, e.g., $\psi(\phi) = \sin(\phi)$ for a monochromatic plane wave field. A constant-crossed field corresponds to a plane-wave field with $\psi(\phi) = \phi$. Locally, a plane-wave field is indistinguishable from a constant-crossed field for any process which happens on a much shorter time scale than $1/\omega_k$ (more details are given below; see also, e.g., \cite{dipiazza_lcfa_2017} and the references therein). 

\subsection{Semiclassical nature of the Volkov states}

For a plane-wave background field the Dirac equation [\eqref{eqn:diracschwingereq} with $M=0$] can be solved exactly and we obtain so-called Volkov states $\Psi(x) \sim \exp{[iS(x)]}$, which have the following phase structure (see, e.g., \cite{landau_quantum_1981}, \S\,40; we focus on electrons in the following)
\begin{gather}
\label{eqn:volkovphase}
S(x) 
=
- p_0x + \int_{-\infty}^{kx} d\phi' \, \lsb \frac{e\, p_0A(\phi')}{\, kp_0} + \frac{e^2A^2(\phi')}{2\, kp_0} \rsb
\end{gather}
($p_0^\mu$ is the asymptotic electron four-momentum outside the field, i.e., at $kx \to -\infty$). Notably, $S(x)$ corresponds to the classical action, i.e., solves the classical Hamilton-Jacobi equation \cite{brown_interaction_1964}
\begin{gather}
[eA^\mu(x) - \del^\mu S(x)] [eA_\mu(x) - \del_\mu S(x)] = m^2.
\end{gather}
Correspondingly, the quantum dynamics inside a plane-wave field is semiclassical. 

For the calculation of probabilities only the relative change of the phase [see \eqref{eqn:volkovphase}] is important, as a total phase drops from physical observables. Therefore, we focus on the change of the action between two space-time points $x^\mu$ and $y^\mu$, which is given by
\begin{gather}
\label{eqn:actionchange}
\Delta S = S(y) - S(x) = \average{p-eA}^\mu(x_\mu-y_\mu).
\end{gather}
Here, we introduced the average (canonical) electron four-momentum between the phases $\phi_x = kx$ and $\phi_y = ky$
\begin{gather}
\label{eqn:averagecanonicalfourmomentum}
\average{p-eA}^\mu = \frac{1}{(\phi_y\!-\!\phi_x)} \int_{\phi_x}^{\phi_y} d\phi' \, [p^\mu(\phi') - eA^\mu(\phi')].
\end{gather}
Furthermore, $p^\mu(\phi)$ denotes the solution of the classical equations of motion (Lorentz force), parametrized via the laser phase $\phi = kx$ instead of the time $t$ [$x^\mu = (t,\spvec{x})$]
\begin{gather}
\label{eqn:classicalcanonicalfourmomentum}
p^\mu(\phi) - eA^\mu(\phi) 
\!=\!
p_0^\mu - k^\mu \lsb \frac{e\, p_0A(\phi)}{kp_0} + \frac{e^2A^2(\phi)}{2\, kp_0} \rsb.
\end{gather}
From \eqref{eqn:actionchange} we conclude that the average (canonical) four-momentum [see \eqref{eqn:averagecanonicalfourmomentum}] plays an important role in quantum calculations.

\subsection{Monochromatic plane wave field}

If the field is monochromatic and one averages the canonical four-momentum [see \eqref{eqn:averagecanonicalfourmomentum}] over a full period [$\Delta\phi = \phi_y - \phi_x = 2\pi$], the famous Brown-Kibble effective momentum is obtained \cite{brown_interaction_1964}
\begin{gather}
\label{eqn:brownkibblemass}
\average{p-eA}^\mu =  p_0^\mu + \frac{M^2_{\text{osc}}}{2kp_0} k^\mu, \quad M^2_{\text{osc}} = e^2\average{-A^2}.
\end{gather}
Correspondingly, the dressed mass is given by $M_{\text{eff}}^2 = \average{p-eA}^2 = m^2 + M^2_{\text{osc}}$. The new scale $M^2_{\text{osc}}$ is independent of $m$ (purely field induced) but also independent of $\hbar$ (completely classical). Obviously, the parameter $\xi = a_0 = \nfrac{M_{\text{osc}}}{m} = \nfrac{eE}{(m\omega_k)}$ is important and light-matter interactions change qualitatively in the so-called ``relativistic regime'' $\xi \gtrsim 1$. In particular, the interaction with the background field becomes non-perturbative and the absorption of up to $\sim \xi^3$ photons from the background field turns out to be relevant \cite{ritus_1985}.

\subsection{Constant-crossed field}

The above definition of the dressed mass $M_{\text{eff}}^2$ [see \eqref{eqn:brownkibblemass}] is only meaningful if the formation region of the considered process is large in comparison with the wavelength of the field, i.e., if $\xi \lesssim 1$ (the formation region consists of the space-time points which give the dominant contribution to the probability amplitude \cite{baier_concept_2005}). In the regime $\xi \gtrsim 1$ (i.e., $M^2_{\text{osc}} \gtrsim m^2$) this assumption is no longer fulfilled, as quantum processes like photon emission and pair production exhibit a phase formation length $\Delta\phi = \phi_y - \phi_x \sim \nfrac{1}{\xi} \ll 2\pi$ \cite{ritus_1985}. Correspondingly, the process happens effectively inside a constant-crossed field and averaging the canonical four-momentum over a full period [see \eqref{eqn:brownkibblemass}] becomes meaningless.

In order to determine the effective four-momentum and the associated dressed mass for processes in a constant-crossed field, we consider the regime $\xi, \chi \gg 1$ in the following. In this regime the phase formation length of typical quantum processes scales as $\Delta\phi = \phi_y - \phi_x \sim \nfrac{\chi^{1/3}}{\xi}$ \cite{baier_quantum_1989} (this scaling will be obtained below from the uncertainty principle, see \secref{sec:plasmafrequency}). Due to the small formation region we can expand the plane-wave four potential around the phase $\phi_z = (\phi_y + \phi_x)/2$
\begin{gather}
A^\mu(\phi) \approx A^\mu(\phi_z) + (\phi - \phi_z) A'^\mu(\phi_z),
\end{gather}
implying that the process happens effectively inside a constant-crossed field [we will employ a gauge where $A^\mu(\phi_z) = 0$]. After averaging the (canonical) electron four-momentum [see \eqref{eqn:averagecanonicalfourmomentum}] over the formation region $\Delta\phi$, we obtain
\begin{gather}
\label{eqn:ccf_dressedmomentum}
\average{p-eA}^\mu =  p^\mu(\phi_z) +  \frac{M_{\text{ccf}}^2}{2kp_0} k^\mu, \,\, M_{\text{ccf}}^2 \sim \Delta\phi^2 \xi^2 m^2.
\end{gather}
Accordingly, $\average{p-eA}^2 \sim M_{\text{ccf}}^2 \sim m^2 \chi^{2/3}$ represents the dressed mass of an electron/positron during a typical quantum transition inside a constant-crossed field.

In order to explain why the field-induced quantum corrections to the electron/positron and the photon mass are of the order of $\mu^2 = \alpha M_{\text{ccf}}^2 = \alpha \chi^{2/3} m^2$ (i.e., contain an additional factor of $\alpha$ with respect to $M_{\text{ccf}}^2$), we have to take into account that they are mediated by the mass [$\mc{M}(x,y)$, see \eqref{eqn:diracschwingereq}] and the polarization operator [$\mc{P}(x,y)$, see \eqref{eqn:photonwaveeq}], respectively. In leading order, both mass and polarization operator require two interactions (see Fig.\,2 of the main text), implying that the probability for the appearance of either loop scales as $\alpha$. 

In summary, the mass scale $\mu^2 = \alpha M_{\text{ccf}}^2$ can be interpreted as the result of an interplay between the dressed mass $M_{\text{ccf}}$ of the fermions in the loops describing quantum fluctuations and the probability $\alpha$ for the appearance of these loops. Below, we will concentrate on photons and provide a more intuitive picture based on the ``plasma frequency of the vacuum'' (see \secref{sec:plasmafrequency}).

\section{``Plasma frequency of the vacuum''}
\label{sec:plasmafrequency}

The wave equation for a photon inside a strong background field [see \eqref{eqn:photonwaveeq}] is both conceptually and mathematically closely related to the propagation of a transverse electromagnetic wave through a cold, collisionless, and nonrelativistic plasma \cite{stix_waves_1992}. In this approximation the electric field of the wave accelerates the electrons of the plasma via the Lorentz force and the generated current emits a wave with the same frequency
\begin{gather}
\label{eqn:plasmawaveeq}
(\del^2_t - \nabla^2) \spvec{A}(x) = \spvec{J}(x), \quad \del_t \spvec{v}(x) = \frac{e}{m} \del_t \spvec{A}(x), 
\end{gather}
where $\spvec{J} = -n_e e \spvec{v}$ and $n_e$ denotes the electron density. Using the plane-wave ansatz $\spvec{A}(x) \sim \spvec{v}(x) \sim \exp{(-iqx)}$ with $q^\mu = (\omega_q, \spvec{q})$, we find the well-known dispersion relation \cite{stix_waves_1992}
\begin{gather}
\label{eqn:plasmadispersion}
\omega_q^2 = \spvec{q}^2 + \Omega_p^2, \quad \Omega_p^2 = \nfrac{n_e e^2}{m},
\end{gather}
where $\Omega_p$ denotes the electron plasma frequency. 

Similarly, a photon propagating through a strong-field vacuum interacts with a ``plasma'' of (virtual) electron-positron pairs, which are dressed by the background field. Therefore, we can interpret the photon mass as the associated plasma frequency [$m_\gamma^2 \sim \alpha \chi^{2/3} m^2$ $\leftrightarrow$ $\Omega_p^2 = \nfrac{n e^2}{m^*}$, see \eqref{eqn:plasmadispersion}]. 

To estimate the density of the (virtual) pair plasma $n$ and the effective mass $m^*$ of the plasma particles we have to identify the space-time region which significantly contributes to the polarization operator, i.e., from which virtual pairs effectively absorb and reemit the propagating photon (i.e., the formation region of the polarization operator). For simplicity, we analyze the case of an electric field with magnitude $E$, oriented transversely to the propagation direction of the photon in the following. We denote the electron, positron and photon four-momentum in the lab frame by $p^\mu = (\eps,p^\perp_1,p^\perp_2,p^\parallel)$, $p'^\mu = (\eps',p'^\perp_1,p'^\perp_2,p'^\parallel)$, and $q^\mu = \omega_q (1,0,0,1)$, respectively, implying $\chi = \chi_q \sim \nfrac{\omega_q Ee}{m^3}$.

At a fundamental QED vertex four-momentum is conserved, i.e., $p^\mu + p'^\mu = q^\mu$. Accordingly, the pair is initially not real ($p^2, p'^2 \neq m^2$). In order to estimate the life time of the virtual transition, we assume the contrary (i.e., $p^2 = p'^2 = m^2$) and determine the energy mismatch. Due to the electric field the momentum changes during the time $\Delta t$ as $p^\perp_1 \to p^\perp_1 - eE \Delta t$. Correspondingly, the energy is given by
\begin{gather}
\label{eqn:epsvsmomentum}
\eps = \sqrt{m^2 + (p^\perp)^2  + (p^\parallel)^2} \approx p^\parallel + \nfrac{(p^\perp)^2}{(2p^\parallel)},
\end{gather}
where $(p^\perp)^2 = (p^\perp_1 - eE\Delta t)^2 + (p^\perp_2)^2 \sim (eE \Delta t)^2$ denotes the transverse momentum. The given expansion is valid if we assume that the field is subcritical in the lab frame ($E \ll \nfrac{m^2}{e}$) but above critical in the boosted frame ($\omega_q \gg m$). As a result, we obtain the hierarchy $(p^\parallel)^2 \gg (p^\perp)^2 \gg m^2$, implying that the rest mass $m$ is negligible from the outset. 

As the conservation law $p^\parallel + p'^\parallel = \omega_q$ has to be valid up to corrections on the scale $p^\perp \sim p'^\perp \sim eE \Delta t$, the energy mismatch is given by [see \eqref{eqn:epsvsmomentum}]
\begin{gather}
\label{eqn:deltaeps}
\Delta\eps = \eps + \eps' - \omega_q \approx  \frac{(p^\perp)^2}{2p^\parallel} + \frac{(p'^\perp)^2}{2p'^\parallel}  \sim \frac{(eE \Delta t)^2}{\omega_q}.
\end{gather}
From the uncertainty principle ($\Delta\eps \, \Delta t  \sim 1$) we find $\Delta t \sim \nfrac{\omega_q}{M^2}$ with $M \sim eE \Delta t \sim m\chi^{1/3}$. Correspondingly, we see that the appearance of the mass scale $M \sim m\chi^{1/3}$ in the limit $\chi \gg 1$ is a consequence of the uncertainty principle, which determines the formation region of the polarization operator. Using $\Delta\phi \sim \omega_k \Delta t$, we recover the scaling of the formation region inside a plane-wave field [see the discussion in \secref{sec:classicalmassdressing}, in particular \eqref{eqn:ccf_dressedmomentum}].

By analyzing the classical equations of motion we find that after the time $\Delta t$ the spatial distance between the electron and the positron is given by $l_\perp \sim \nfrac{1}{M}$ (transversely) and $l_\parallel \sim (v_\parallel-v'_\parallel) \Delta t \sim \nfrac{1}{\omega_q}$ (longitudinal); the last relation follows from $1 - v_\parallel = 1 - (\nfrac{p^\parallel}{\eps}) \sim (\nfrac{M}{\omega_q})^2$. As the annihilation of the pair is only possible if its spatial distance is within the volume allowed by the uncertainty principle, we can interpret $V \sim l_\parallel l_\perp^2 \sim \nfrac{1}{(\omega_q M^2)}$ as the space-time volume which the pair is allowed to occupy (a similar spreading argument explains the recollision probability of the pair \cite{meuren_high-energy_2015}). 

In order to relate the volume $V \sim \nfrac{1}{(\omega_q M^2)}$ to the plasma density, we note that the derivation of the dispersion relation above [see \eqref{eqn:plasmadispersion}] is only valid in the plasma rest frame. Due to the contraction ($l_\parallel \ll l_\perp$) of the spatial volume $V \sim l_\parallel l_\perp^2$, the laboratory frame cannot correspond to the effective rest frame of the virtual pair plasma. From length contraction $l_\parallel = \nfrac{\tilde{l}_\parallel}{\tilde{\gamma}}$ we conclude that we need a Lorentz boost with gamma factor $\tilde{\gamma} \sim \nfrac{\omega_q}{M}$ along the direction of the momentum of the incoming photon to obtain a uniform volume, i.e., $\tilde{l}_\parallel \sim \tilde{l}_\perp \sim \nfrac{1}{M}$. In this frame the electron/positron has an energy $\tilde{\eps} \sim \tilde{\eps}' \sim M$. If we interpret this frame as the effective rest frame of the pair, we should assign the pair the effective mass $m^* \sim M$, in agreement with the result obtained more rigorously above by considering the dressed mass of an electron/positron [see \eqref{eqn:ccf_dressedmomentum}]. 

Finally, we find that in the effective rest frame of the pair the volume is given by $\tilde{V} \sim \tilde{l}_\parallel \tilde{l}_\perp^2 \sim \nfrac{1}{M^3}$, implying $n \sim M^3$. Using $m^* \sim M$ and $\Omega_p^2 = \nfrac{n e^2}{m^*}$ we obtain that the ``plasma frequency of the vacuum'' (i.e., the photon mass) is given by $\Omega_p^2 \sim \alpha \chi^{2/3} m^2$, in agreement with the exact evaluation of the polarization operator [see \eqref{eqn:ccfphotonmasses}] \cite{narozhnyi_propagation_1968}.